\documentclass[10pt,aps,prd,showpacs,superscriptaddress,twocolumn]{revtex4}
\usepackage{amsfonts}
\usepackage{amssymb}
\usepackage{amsmath}
\usepackage{MnSymbol}
\usepackage{graphicx}

\date{\today}

\begin{document}

\title{Nonperturbative approach to relativistic quantum communication channels}

\author{Andr\'e G.\ S.\ Landulfo}\email{andre.landulfo@ufabc.edu.br}
\affiliation{Centro de Ci\^encias Naturais e Humanas,
Universidade Federal do ABC, 
Avenida dos Estados, 5001, 09210-580, Bangu,
Santo Andr\'e, S\~ao Paulo, Brazil}

\begin{abstract}

We investigate the transmission of both classical and quantum information between two arbitrary observers in globally hyperbolic spacetimes using a quantum field as a communication channel. The field is supposed to be in some arbitrary quasifree state and no choice of representation of its canonical commutation relations is made. Both sender and receiver possess some localized two-level quantum system with which they can interact with the quantum field to prepare the input and receive the output of the channel, respectively. The interaction between the two-level systems and the quantum field is such  that one can trace out the field degrees of freedom exactly and thus obtain the quantum channel in a nonperturbative way. We end the paper determining the unassisted as well as the entanglement-assisted classical and quantum channel capacities.
\end{abstract}

\pacs{04.62.+v, 03.67.-a, 03.67.Hk}

\maketitle

\section{Introduction}
\label{sec:introduction}

How much classical and quantum information can be conveyed between two (or more) parts when one uses a quantum communication channel? This is one of the fundamental questions addressed by quantum information theory~\cite{wilde, GT07} and it is within such quantum communication scenarios  that relativity can play a major role. This will be the case if, e.g., the sender and the receiver are set in a  relativistic relative motion or if the spacetime possesses some nontrivial structures such as Cauchy or event horizons~\cite{wald84}. 

In order to analyze quantum information theory in general spacetimes and in a relativistically consistent way, one should use quantum field theory in curved spacetimes~\cite{wald94}. This was done by several authors to study, e.g., the behavior of entanglement with respect to different observers~\cite{FM05,MM10, FMMM10,ML10,MGL10,BLMDF10, MF11,RO15} or the transmission of classical and quantum information~\cite{AM03, BHTW10, MHM12, BHP12, HBK12, LT13, BA15}. Most of the analysis was done in the context of Minkowski, Schwarzschild, or asymptotically flat cosmological spacetimes. In all such spacetimes there are at least two congruences of observers covering some open set and following the orbits of a timelike Killing field. Thus, one is able to naturally choose two different Fock space representations of the canonical commutation (or anticommutation) relations of the field, i.e., two different notions of ``particles,"  one associated with each congruence of observers. As a consequence, one can analyze the communication between these two set of observers or the entanglement with respect to each of them.

It would be interesting however, to generalize these quantum communications scenarios to describe information transfer in more general spacetimes, in a more ``covariant" manner, i.e., without the need for any particular choice of particles, and allowing more general observers conveying the information, not being restricted to the ones following the orbits of some Killing field (which may not exist given a general spacetime).   Additionally, there is another aspect that would be interesting to have incorporated in such relativistic communication models: Typically, both the sender and receiver of the information can measure the state of the field being used for communication in a limited region of spacetime. We can imagine them as experimentalists restricted to their laboratories and doing their experiments during a finite proper time. Communication using quantum fields but with the sender and receiver having access to a bounded region of spacetime have drawn much attention recently. By using Unruh-DeWitt detectors~\cite{dewitt} to model the local interaction between each of the parts that are communicating and the quantum field, it was possible to do a perturbative investigation of the classical communication capacity  between inertial observers in Minkowski spacetime~\cite{CK10}, of quantum signaling in cavity QED~\cite{JMK14}, and of the possibility of timelike information transfer in cosmological spacetimes~\cite{MM} (for a detailed analysis of causality issues associated with inertial Unruh-DeWitt detectors in Minkowski spacetime see~\cite{MM15}). In Refs.~\cite{H, HL,YH10,MM2}, by using harmonic oscillators instead of Unruh-DeWitt detectors, it was possible to do a nonperturbative analysis of the response of the detectors, teleportation, and entanglement harvesting and sudden death in this setup. 

In the present paper we will describe a model of communication using bosonic quantum fields in arbitrary globally hyperbolic spacetimes, without choosing any particular Fock space representation of the canonical commutation relations, and allowing an arbitrary motion for both sender and receiver. The two observers will use two-level quantum systems (qubits) to locally interact with the quantum field and convey the information. The field will be in some arbitrary quasifree state~\cite{wald94} and the interactions between each of the two-level systems and the field will be very similar to the ones given by the Unruh-DeWitt model. We will suppose,  however, that the two levels of the qubit have the same energy and hence, in the absence of noise, the state prepared for the qubit does not change. This model is interesting because, as it will be shown during the paper, one can trace out the field degrees of freedom exactly and hence, study the quantum channel defined between the sender and the receiver nonperturbatively.  With the mathematical description of the quantum channel we will be able to determine the rate at which classical and quantum information can be reliably transmitted  (i.e. with arbitrarily small error probability) in such a communication channel. In particular, the causality in the information transmission will be manifest.

The paper is organized as follows. In Sec.~\ref{sec:quantization} we will describe the quantization of a free scalar field on globally hyperbolic spacetimes as well as the class of states we will be using. In Sec.~\ref{sec:channel}, we will describe the interaction between the qubits and the field and determine the quantum map that describes the communication channel. In Sec.~\ref{sec:capacities}, we will use the quantum map derived in Sec. III to calculate various channel capacities. Section~\ref{sec:finalremarks} is reserved to our final remarks. We assume metric signature $(- + + +)$ and natural units in which $c=\hbar=G=1$ unless stated otherwise.

\section{Field quantization}
\label{sec:quantization}

Let us consider a free, real scalar field $\phi$ propagating in a four-dimensional globally hyperbolic spacetime $(\mathcal{M},g_{ab})$.  Let the spacetime be foliated by Cauchy surfaces  $\Sigma_t$ labeled by the real parameter $t$. The field is described by the action
\begin{equation}\label{kg_action}
S \equiv 
-\frac{1}{2}
\int_{\mathcal{M}} d^4x \sqrt{-g} \,
( \nabla_a\phi \nabla^a\phi + {\rm m}^2 \phi^2 + \xi R \phi^2 ),
\end{equation}
which gives rise to the Klein-Gordon equation
\begin{equation}
(-\nabla^a\nabla_a + {\rm m}^2 +\xi R)\phi =0.
\label{KG}
\end{equation}
Here ${\rm m}$ is the field mass, $\xi\in \mathbb{R}$, $R$ is the scalar curvature, $\nabla_a$ is the torsion-free  covariant derivative compatible with $g_{ab}$, and $g \equiv {\rm det} (g_{\mu\nu})$ in some arbitrary coordinate system. 

We will use the canonical quantization procedure, which consists of promoting the field to an operator (rigorously, an operator-valued distribution) that satisfies the canonical commutation relations (CCR)
\begin{eqnarray}
\, [\phi (t, {\bf x}), \phi (t, {\bf x}') ]_{\Sigma_t} 
& = & 
[\pi  (t, {\bf x}), \pi  (t, {\bf x}') ]_{\Sigma_t} = 0,
\label{canonical_quantization1} 
\\
\, [ \phi (t, {\bf x}), \pi  (t, {\bf x}')]_{\Sigma_t}
& = &  
i \delta^3 ({\bf x}, {\bf x}' ),
\label{canonical_quantization2}
\end{eqnarray}
where ${\bf x} \equiv (x^1,x^2,x^3)$ are coordinates on the Cauchy surface $\Sigma_t$. The conjugate momentum $\pi(x)$ is defined as 
\begin{equation}\label{conjugate_canonical_momentum}
\pi\equiv {\delta S}/{\delta \dot{\phi}} = \sqrt{^{(3)} g\, }\, n^a\nabla_a\phi,
\end{equation}
where $n^a$ is the future-directed unit vector field normal to $\Sigma_t$, $``\; \dot{}\;" \equiv \partial_t $, and $^{(3)} g \equiv {\rm det} (^{(3)}g_{i j})$, with $^{(3)} g_{ij}$ being the induced metric on $\Sigma_t$ written in the coordinates $(x^1,x^2,x^3)$. 

We may formally write the time evolution of the field operator $\phi$, with respect to the foliation $\Sigma_t$, as $\phi(t,{\bf x})=U_\phi^\dagger(t)\phi(0,{\bf x})U_\phi(t)$ with $U_\phi(t)$ satisfying
\begin{equation}
\dot{U}_\phi(t)=-iH_\phi(t)U_\phi (t),
\end{equation}
where 
\begin{equation}
H_\phi (t)\equiv \int_{\Sigma_t}d^3{\bf x}( \pi(t,{\bf x})\dot{\phi}(t,{\bf x}) - \mathcal{L}[\phi,\nabla_a\phi])
\label{canonicalH}
\end{equation}
is the canonical Hamiltonian with 
$$\mathcal{L}[\phi,\nabla_a\phi]\equiv-\frac{1}{2}\sqrt{-g} \,( \nabla_a\phi \nabla^a\phi + {\rm m}^2 \phi^2 + \xi R \phi^2 )$$
being the Lagrangian density.

In order to find a representation of the CCR, Eqs.~(\ref{canonical_quantization1}) and  (\ref{canonical_quantization2}), we will first define the antisymmetric bilinear map $\Omega$ on a space $\mathcal{S}^\mathbb{C}$ of complex solutions of Eq.~(\ref{KG}) as
\begin{equation}
\Omega(\varphi_1,\varphi_2)
\equiv 
\int_{\Sigma_t} d\Sigma \, n^a [\varphi_2 \nabla_a \varphi_1 - \varphi_1 \nabla_a \varphi_2 ],
\label{OmegaKG}
\end{equation}
where $d\Sigma$ stands for the proper-volume element on $\Sigma_t$. We recall that $\Omega(\varphi_1,\varphi_2)$ do not depend on  the Cauchy surface $\Sigma_t$. With Eq.~(\ref{OmegaKG}), we can define the Klein-Gordon inner product as
\begin{equation} 
(\varphi_1,\varphi_2)_{\rm KG}\equiv -i\Omega(\overline{\varphi}_1,\varphi_2),
\label{inner_KG}
\end{equation}
and we note that the above inner product is not positive definite on $\mathcal{S}^\mathbb{C}$. Now, we choose any subspace $\mathcal{H}\subset \mathcal{S}^\mathbb{C}$ such that ({\bf i}) $\mathcal{S}^\mathbb{C}\simeq\mathcal{H}\oplus\overline{\mathcal{H}}$; ({\bf ii}) the inner product~(\ref{inner_KG}) is positive definite in $\mathcal{H}$ and  makes $\left(\mathcal{H}, (,)_{\rm KG}\right)$ a Hilbert space; and ({\bf iii}) given any $u\in \mathcal{H}$ and $v\in\mathcal{\overline{H}}$, $(u,v)_{\rm KG}=0$~\cite{footnote1}. We can now define the field's Hilbert space as the symmetric Fock space $\mathfrak{F}_{\rm s}(\mathcal{H})$ and  the field operator as 
\begin{equation}
\phi(t,{\bf x})\equiv \sum_j \left[u_j(t,{\bf x}) a(\overline{u}_j)+\overline{u}_j(t,{\bf x}) a^\dagger(u_j)\right], 
\label{field_op1}
\end{equation}
where  $\{u_j\}$ is an orthonormal basis for $\mathcal{H}$ and $a(\overline{u})$ and $a^\dagger(v)$, satisfying
\begin{equation}
[a(\overline{u}),a^\dagger(v)]=(u,v)_{\rm KG}I,
\label{ca_ccr}
\end{equation}
with $I$ being the identity operator, are the usual annihilation and creation operators of modes $u$ and $v$, respectively, where $u,v\in \mathcal{H}$. With the field operator defined in Eq.~(\ref{field_op1}) we get a representation of the CCR on $\mathfrak{F}_s(\mathcal{H})$. The vacuum state associated with this representation is the normalized vector $|0\rangle\in\mathfrak{F}_s(\mathcal{H})$ which satisfies $a(\overline{u})|0\rangle=0$ for all $u\in\mathcal{H}$.

To be mathematically well defined, the field operator~(\ref{field_op1}) must be seen as an operator-valued distribution. It is not difficult to see that if we smear out $\phi(x)$ in Eq.~(\ref{field_op1}) with a test function $f\in C^\infty_0(\mathcal{M})$, where $C^\infty_0(\mathcal{M})$ stands for the set of all smooth, compact-support real functions on $\mathcal{M}$, we get~\cite{wald94}
\begin{equation}
\phi(f) = i\left[a(\overline{KEf})-a^\dagger(KEf)\right]. 
\label{field_op2}
\end{equation}
Here, the operator $K:\mathcal{S}\rightarrow \mathcal{H}$ takes the positive-norm part of any real solution $\varphi\in\mathcal{S}$, with $\mathcal{S}\subset \mathcal{S}^\mathbb{C}$ being a space of real solutions of Eq.~(\ref{KG}),  and, for any  $f\in C^\infty_0(\mathcal{M})$, 
\begin{equation}
Ef(x) \!\equiv \!\int_{\mathcal{M}} d^4x'\sqrt{-g(x')} [G^{\rm adv}(x, x')-G^{\rm ret}(x, x')] f(x'),
\label{Ef}
\end{equation}
where $G^{\rm adv}$ and $ G^{\rm ret}$ are the advanced and retarded Green functions of the operator $P\equiv -\nabla^a\nabla_a + {\rm m}^2 +\xi R$, respectively. We note that, given any test function $f$, $Ef$ is a solution of Eq.~(\ref{KG}). Using Eqs.~(\ref{ca_ccr}) and~(\ref{field_op2}) we get the covariant version of the CCR:
\begin{equation}
[\phi(f_1),\phi(f_2)]=-i\Delta(f_1,f_2)I,
\label{CCR2}
\end{equation}
where
\begin{equation}
\Delta(f_1,f_2)\equiv\int_{\mathcal{M}} d^4x\sqrt{-g}f_1(x)Ef_2(x)
\label{CCR3}
\end{equation}
and $f_1,f_2\in C^\infty_0(\mathcal{M})$.  

The above construction is not unique; there are infinitely many choices of $\mathcal{H}$ satisfying conditions ({\bf i})-({\bf iii}) defined below Eq.~(\ref{inner_KG}). Hence there are infinitely many Fock space representations of the CCR and thus, infinitely many choices of vacuum and particles.  As a general rule, the different representations are not unitarily equivalent. Contrary to what this may suggest, this poses no difficulty to the formulation of the theory. We can see this many inequivalent representations of the CCR in quantum field theory in curved spacetimes as analogous to coordinates in general relativity~\cite{wald94}. This is made manifest when one uses the algebraic approach to quantum field theory~\cite{wald84, FB}. In this formulation, the field quantization can be seen as a $\mathbb{R}$-linear map $\phi: f\in C^\infty_0(\mathcal{M})\rightarrow \phi(f)\in\mathcal{A}(\mathcal{M})$ between $C^\infty_0(\mathcal{M})$ and a $^*$-algebra $\mathcal{A}(\mathcal{M})$, which satisfies: 
\begin{enumerate}
\item{$\phi(f)^{*}=\phi(f)$, $f\in  C^\infty_0(\mathcal{M})$, i.e., the (smeared) field is Hermitian,}
\item{ $\phi([-\nabla^a\nabla_a + {\rm m}^2 +\xi R]f)=0$, for all $f\in  C^\infty_0(\mathcal{M})$, i.e., the field satisfies the Klein-Gordon equation,}
\item{$\left[\phi(f_1),\phi(f_2)\right]=-i\Delta(f_1,f_2)I,$ $f_1, f_2\in  C^\infty_0(\mathcal{M})$, i.e., the field satisfies the CCR,}
\item{$\mathcal{A}(\mathcal{M})$ is algebraically generated by the identity $I$ and the $\phi(f)$'s, $f\in  C^\infty_0(\mathcal{M})$.}
\end{enumerate}
We call $\mathcal{A}(\mathcal{M})$ the algebra of observables of the Klein-Gordon field.

A quantum state is a $\mathbb{C}-$linear functional $\omega : \mathcal{A}(\mathcal{M}) \rightarrow \mathbb{C}$ which satisfies  (S1) positivity, i.e., $\omega(A^*A)\geq 0$ for all $A\in \mathcal{A}(\mathcal{M})$ and (S2) normalization, i.e., $\omega(I)=1.$ Given a $\lambda\in (0,1)$ and any two states $\omega_1$ and $\omega_2$, we note that $\lambda \omega_1 + (1-\lambda)\omega_2$ also defines a state. We say that a quantum state $\omega$ is pure if and only if every decomposition of $\omega$ as $\omega=\lambda \omega_1 +(1-\lambda)\omega_2$ implies that $\omega_1=\omega_2=\omega$, otherwise, we say that $\omega$ is a mixed state.  By using the so-called Gelfand-Naimark-Segal (GNS construction~\cite{wald94,FB}, it can be shown that every state $\omega$ can be realized as a vector on a Hilbert space $\mathfrak{H}$ carrying a  representation of the algebra of observables. The advantage of the algebraic formulation however, is that, although it allows us to choose a representation of the CCR, we need not do so. The algebraic approach enables us to formulate quantum field theory in curved spacetimes without forcing us to make  arbitrary choices. 

For our purposes it will be interesting to work not with $\mathcal{A}(\mathcal{M})$ but with its ``exponentiated version"  $\mathcal{W}(\mathcal{M})$, since it will be its elements that will appear in our description of the quantum communication channel. The algebra $\mathcal{W}(\mathcal{M})$ is called the Weyl algebra and  it is the $^*$-algebra~\cite{footnote2} generated by the elements 
\begin{equation}
W(Ef)\equiv e^{i\phi(f)}, f\in C^\infty_0(\mathcal{M}),
\end{equation} 
satisfying:
\begin{eqnarray}
&& W(Ef)^*=W(-Ef), \label{W1}\\
&& W(E[-\nabla^a\nabla_a + {\rm m}^2 +\xi R]f)=I, \label{W2}\\
&& W(Ef_1)W(Ef_2)=e^{\frac{i}{2}\Delta(f_1,f_2)}W[E(f_1+f_2)], \label{W3}
\end{eqnarray}  
with $f, f_1,f_2\in  C^\infty_0(\mathcal{M})$. Equations~(\ref{W1})-(\ref{W3}) are the exponentiated version of  conditions 1-3 defined earlier. An algebraic state  is then defined as a positive and normalized $\mathbb{C}$-linear functional $\omega:\mathcal{W}(\mathcal{M})\rightarrow \mathbb{C}.$ 

There is one class of states that stands out from the plethora of algebraic states, the so-called quasifree states. In order to define this class of states, let us first define $\mathcal{S}\equiv \{Ef |f\in C^\infty_0(\mathcal{M}) \}$. As a real vector space, this set coincides  with the space of all real solutions of Eq.~(\ref{KG}) with compact-support initial data on $\Sigma_t$~\cite{wald94, FB}. Now, given a  real inner product $\mu:\mathcal{S}\times \mathcal{S}\rightarrow \mathbb{R}$ on $\mathcal{S}$ satisfying
\begin{equation}
\left|\Omega(Ef_1,Ef_2)\right|^2 \leq 4 \mu(Ef_1,Ef_1)\mu(Ef_2,Ef_2),
\end{equation}
we define a quasifree state $\omega_\mu$, associated with $\mu$, by the relation
\begin{equation}
\omega_\mu[W(Ef)] \equiv e^{-\mu(Ef,Ef)/2},
\label{quasifree}
\end{equation}
for all $f\in  C^\infty_0(\mathcal{M})$, where the action of $\omega_\mu$ on an arbitrary element of $\mathcal{W}(\mathcal{M})$ is defined by linearity and continuity. The vacuum states that are usually considered in quantum field theory in curved spacetimes are pure quasifree states~\cite{KayWald91} but we will not restrict ourselves to this case. Allowing the quasifree states to be statistical mixtures will enable us to consider several important classes of states such as, for instance, the thermal equilibrium [i.e. Kubo-Martin-Schwinger (KMS)] states. 


\section{The quantum channel}
\label{sec:channel}

Suppose now that two observers, Alice and Bob, want to communicate using the quantum field $\phi$ as a communication channel. We will consider that the field is in some quasifree state $\omega_\mu$.  It should be noted however that, although we are focusing on this class of states, the derivation we will present is valid for any algebraic state $\omega$ which satisfies 
\begin{equation}
\omega[W(Ef)]\in\mathbb{R}_+,
\label{generalstate}
\end{equation} 
for all $f\in C^\infty_0(\mathcal{M})$. This includes not only the quasifree states but, e.g.,  also n-particle states in any given representation of the canonical commutation relations. Each observer possesses a two-level gapless quantum system with which they can interact with the quantum field. We denote the two-dimensional Hilbert spaces associated with Alice's and Bob's qubits by $\mathcal{H}_A$ and $\mathcal{H}_B$, respectively. In order to convey classical or quantum information to Bob, Alice prepares her qubit in some quantum state $\rho^A_{-\infty}$ and turns on its interaction with the field for a finite time (with respect to the foliation $\Sigma_t$) $\Delta t_A.$ After that, Bob switches on his qubit  interaction with the field for a time interval $\Delta t_B$ in order to measure the information imprinted by Alice on the field's state. Bob's qubit is initially prepared in a suitable ready-to-measure state $\rho^B_{-\infty}$.

The total Hamiltonian of the two-qubit$+$field system is given by
\begin{equation}
H\equiv H_\phi + H_{\rm int},
\label{totalH}
\end{equation}
where $H_\phi$ is given by Eq.~(\ref{canonicalH}) and $H_{\rm int}$ is the interaction Hamiltonian which, in the interaction picture, is given by
\begin{equation}
H_{\rm int}^I \equiv \sum_{j}\epsilon_j(t)\int_{\Sigma_t}d^3{\bf x} \sqrt{-g} \psi_j(t,{\bf x}) \phi(t,{\bf x}) \otimes\sigma^{\rm z}_j.
\label{Hint}
\end{equation} 
Here, $j=A, B$, with $A$ and $B$ labeling Alice's and Bob's qubit, respectively;  $\sigma^{\rm z}_j$ is one of the  Pauli matrices $\sigma^{\rm x}_j,\sigma^{\rm y}_j,\sigma^{\rm z}_j$ associated with qubit $j$; $\epsilon_j \in C^\infty_0(\mathbb{R})$ is a time-dependent coupling constant which keeps the interaction of qubit $j$ with the field switched on for a finite time  $\Delta t_j$; and  $\psi_j(t,{\bf x})$ is a smooth real function satisfying  $\psi_j|_{\Sigma_t}\in C^\infty_0(\Sigma_t)$ for all $t$, which models the fact that the qubit $j$ interacts with the field only in a vicinity of its worldline.  We note that, for the communication protocol chosen, the support of each coupling constant is given by $${\rm supp}\; \epsilon_j=[T_j^i,T_j^f],$$ with $T_j^f-T_j^i=\Delta t_j$ and $T_B^i\geq T_A^f,$ i.e., the interaction of qubit B with the field is switched on after qubit A's interaction is  switched off.

The interaction picture time evolution operator, associated with the foliation $\Sigma_t$, can be written as
 \begin{equation}
U=T\exp{\left[-i\int_{-\infty}^\infty dt H^I_{\rm int}(t)\right]}, 
\label{unitaryevo}
\end{equation} 
where $T$ indicates time ordering with respect to $t$. By using the Magnus expansion~\cite{magnus}
\begin{equation}
\Omega\equiv\sum_{n=1}^\infty\Omega_n,
\label{magnus1}
\end{equation}
where each $\Omega_n$ is an operator of order $n$ in $H^I_{\rm int}(t)$, we can cast Eq.~(\ref{unitaryevo}) as
\begin{equation}
U=\exp{\Omega}. 
\label{magnus2}
\end{equation}
The first terms of the expansion~(\ref{magnus1}) are given by 
\begin{eqnarray}
\Omega_1&=&-i\int_{-\infty}^\infty dt H^I_{\rm int}(t), \\
\Omega_2&=&-\frac{1}{2}\int_{-\infty}^\infty dt\int_{-\infty}^{t} dt'[ H^I_{\rm int}(t),H^I_{\rm int}(t')] \\
\Omega_3&=&\frac{i}{6}\int_{-\infty}^\infty \!\!\!\!\!\!dt \!\!\!\int_{-\infty}^{t}\!\!\!\!\!dt'\!\!\!\int_{-\infty}^{t'}\!\!\!\!\!\!dt'' \left([H^I_{\rm int}(t) ,[H^I_{\rm int}(t'),H^I_{\rm int}(t'')] ] \right. \nonumber \\ 
&+& \left.[H^I_{\rm int}(t'') ,[H^I_{\rm int}(t'),H^I_{\rm int}(t)] ]\right),
\label{omega}
\end{eqnarray}
and the high-order terms can be obtained recursively. Using the explicit form of the interaction Hamiltonian~(\ref{Hint}) and the covariant canonical commutation relation~(\ref{CCR2}), it can be seen that 
\begin{eqnarray}
\Omega_1&=&-i\sum_j\phi(f_j)\otimes  \sigma^{\rm z}_j, \label{Omega1}\\
\Omega_2&=& i \; \Xi \; I - \frac{i}{2} \Delta(f_A,f_B)\sigma^{\rm z}_A\otimes\sigma^{\rm z}_B,
\label{Omega2}
\end{eqnarray} 
and 
\begin{equation}
\Omega_k=0, \; \mathrm{for}\; k\geq 3. 
\label{Omegak}
\end{equation}
Here,
\begin{equation}
f_j(t,{\bf x})\equiv \epsilon_j(t) \psi_j(t,{\bf x})
\label{f}
\end{equation}
is a compact-support function on $\mathcal{M}$ carrying the information about qubit $j$,
\begin{eqnarray*}
\Xi \equiv \frac{1}{2}\sum_j\int_{-\infty}^\infty dt \; \epsilon_j(t)\int_{-\infty}^{t} \; dt' \epsilon_j(t') \Delta_{j}(t,t')
\end{eqnarray*}
with $$\!\Delta_{j}(t,t')\!\equiv \!\!\!\int_{\Sigma_t}\!\!\!\!\!\!d^3{\bf x}\sqrt{-g} \!\!\int_{\Sigma_{t'}}\!\!\!\!\!\!\!d^3{\bf x'} \sqrt{-g'}\psi_j(t,{\bf x})\Delta(x,x')\psi_j(t',{\bf x'}),$$
and we recall that $\phi(f_j)\equiv \int_{\mathcal{M}}d^4x\sqrt{-g}f_j(x)\phi(x)$  and $[\phi(x),\phi(x')]\equiv-i\Delta(x,x')I$ [which is the unsmeared version of Eq.~(\ref{CCR2})]. We note that, in order to obtain the second term in Eq.~(\ref{Omega2}), we have used that $T_B^i\geq T_A^f$. By making use of Eqs.~(\ref{magnus1}) and~(\ref{Omega1})-(\ref{Omegak}) in Eq.~(\ref{magnus2}), as well as the  Zassenhaus formula
$$e^{\mathfrak{a}+\mathfrak{b}}=e^{\mathfrak{a}}e^{\mathfrak{b}}e^{-\frac{1}{2}[\mathfrak{a},\mathfrak{b}]},$$
where $[\mathfrak{a},\mathfrak{b}]$ is a c-number, we get the following expression for the unitary evolution of the system:
\begin{equation}
U=e^{i\Xi}e^{-i\phi(f_A)\otimes \sigma^{\rm z}_A}e^{-i\phi(f_B)\otimes \sigma^{\rm z}_B}e^{-i\Delta(f_A,f_B)\sigma^{\rm z}_A\otimes\sigma^{\rm z}_B}.
\label{unitaryfinal}
\end{equation}
For what follows, it will be useful to cast the unitary operator $\exp{[-i\phi(f_j)\otimes\sigma^{\rm z}_j]}$ as 
\begin{equation}
e^{-i\phi(f_j)\otimes \sigma^{\rm z}_j}=\cos{[\phi(f_j)]}-i\sin{[\phi(f_j)]}\otimes \sigma^{\rm z}_j,
\end{equation}
with 
\begin{equation}
\cos{[\phi(f_j)]}\equiv \frac{1}{2}\left[W(Ef_j)+ W(-Ef_j)\right]
\label{cos}
\end{equation}
and
\begin{equation}
\sin{[\phi(f_j)]}\equiv \frac{1}{2i}\left[W(Ef_j)- W(-Ef_j)\right]
\label{sin}
\end{equation}
being elements of the Weyl algebra $\mathcal{W}(\mathcal{M})$. 

Now, by using the unitary operator given in Eq.~(\ref{unitaryfinal}), we can formally evolve the initial state of the two-qubit$+$field system and then trace out the field degrees of freedom. By doing this, we obtain the state describing the qubits after the communication protocol has ended. The initial state of the qubits is given by $\rho^A_{-\infty}\otimes\rho^B_{-\infty}$ and we recall that the field is in some quasifree state $\omega_{\mu}$ [or, more generally, in some algebraic state satisfying Eq.~(\ref{generalstate})].  Hence, 
\begin{eqnarray}
\rho^{AB}&\equiv& {\rm tr}_\phi \left(U \rho^A_{-\infty}\otimes\rho^B_{-\infty}\otimes \rho_\omega U^\dagger \right)\nonumber \\
&=& \Gamma_{cccc}\tilde{\rho}^{AB} +  \Gamma_{ssss}\sigma^{\rm z}_A\otimes\sigma^{\rm z}_B\tilde{\rho}^{AB}\sigma^{\rm z}_A\otimes\sigma^{\rm z}_B\nonumber \\
&+& \Gamma_{cssc}\sigma^{\rm z}_A\tilde{\rho}^{AB}\sigma^{\rm z}_A+ \Gamma_{sccs}\sigma^{\rm z}_B\tilde{\rho}^{AB}\sigma^{\rm z}_B\nonumber \\
&+&(\Gamma_{cscs}\sigma_B^{\rm z}\tilde{\rho}^{AB}\sigma_A^{\rm z}+ \; {\rm H.c.}) \nonumber \\
&-&(\Gamma_{sscc}\tilde{\rho}^{AB}\sigma_A^{\rm z}\otimes\sigma_B^{\rm z}+ \; {\rm H.c.}), 
\label{rhoAB}
\end{eqnarray}
where 
\begin{equation}
\tilde{\rho}^{AB}\equiv e^{-i\Delta(f_A,f_B)\sigma^{\rm z}_A\otimes\sigma^{\rm z}_B} \left(\rho^A_{-\infty}\otimes\rho^B_{-\infty}\right)e^{i\Delta(f_A,f_B)\sigma^{\rm z}_A\otimes\sigma^{\rm z}_B}
\label{rhotilde}
\end{equation}
and 
\begin{equation}
\Gamma_{\alpha\beta\gamma\delta}\!\equiv \!\omega_{\mu}\!\left(\!\mathcal{F}_\alpha[\phi(f_B)]\mathcal{F}_\beta[\phi(f_A)]\mathcal{F}_\gamma[\phi(f_A)]\mathcal{F}_\delta[\phi(f_B)]\right),\\
\label{Gamma}
\end{equation}
with $\alpha,\beta,\gamma,\delta\in \{c,s\}$, $\mathcal{F}_s(x)\equiv\sin{x}$, and $\mathcal{F}_c(x)\equiv\cos{x}$. In order to do the time evolution in the first line of Eq.~(\ref{rhoAB}), we have written the algebraic $\omega_\mu$ as if it were a density matrix $\rho_\omega$ with ${\rm tr}(\rho_\omega W[Ef])\equiv\omega_\mu(W[Ef])$. Note however that $\rho^{AB}$ (as well as the quantum map between qubits $A$ and $B$ that will be defined later) depends only on the action of $\omega_\mu$ on the Weyl algebra $\mathcal{W}(\mathcal{M})$.  A direct calculation using Eqs.~(\ref{quasifree}),~(\ref{cos}), and~(\ref{sin}) in Eq.~(\ref{Gamma}), as well as Eq.~(\ref{W3}) to simplify products of operators $W(Ef_j)$,  shows that \begin{eqnarray*}
\Gamma_{cccc}\!\!\!&=&\!\!\frac{1}{8}(\nu^+_{AB}+\nu^-_{AB})+\frac{1}{4}(1+\nu_B + \nu_A\cos{[2\Delta(f_A,f_B)]}), \\
\Gamma_{ssss}\!\!\!&=&\!\!\frac{1}{8}(\nu^+_{AB}+\nu^-_{AB})+\frac{1}{4}(1-\nu_B - \nu_A\cos{[2\Delta(f_A,f_B)]}),  \\
\Gamma_{sccs}\!\!\!&=&\!\!\!-\frac{1}{8}(\nu^+_{AB}\!+\nu^-_{AB})\!+\!\frac{1}{4}(1-\nu_B + \nu_A\cos{[2\Delta(f_A,f_B)]}),  \\
\Gamma_{cssc}\!\!\!&=&\!\!\!-\frac{1}{8}(\nu^+_{AB}\!+\nu^-_{AB})\!+\!\frac{1}{4}(1+\nu_B - \nu_A\cos{[2\Delta(f_A,f_B)]}),  \\
\Gamma_{sscc}\!\!\!&=&-\frac{1}{8}(\nu^+_{AB}-\nu^-_{AB}) +\frac{i}{4}\nu_A\sin{[2\Delta(f_A,f_B)]},\\
\Gamma_{cscs}\!\!\!&=&\overline{\Gamma}_{sscc},
\label{Gamma2}
\end{eqnarray*}
with the remaining $\Gamma_{\alpha\beta\gamma\delta}$ vanishing. In the above equations we have used that 
\begin{eqnarray}
\nu_j&\!\!\equiv& \!\!\omega_\mu\left(W[E(2f_j)]\right)=\exp{(-2\|Ef_j\|^2)},  \\
\nu_{AB}^\pm \!\! &\equiv& \!\! \omega_\mu\left(W[E(2f_A\pm2f_B)]\right)=\exp{(-2\|E(f_A\pm f_B)\|^2)}, \nonumber \\
\label{nu}
\end{eqnarray}
with $\|Ef\|^2\equiv \mu(Ef,Ef), f\in C^\infty_0(\mathcal{M}).$
To find the state of Bob's qubit after the protocol has ended, we need to take the partial trace over qubit $A$,
\begin{equation}
\rho^B\equiv {\rm tr}_A \rho^{AB}.
\label{rhoB1}
\end{equation}
By using Eqs.~(\ref{rhoAB}) and~(\ref{rhotilde}), as well as the explicit expressions for $\Gamma_{\alpha\beta\gamma\delta}$, we find that
\begin{eqnarray}
\rho^B&=&\frac{1}{2}\left(1+ \nu_B \cos{[2\Delta(f_A,f_B)]}\right)\rho^B_{-\infty}, \nonumber \\
&+& \frac{1}{2}\left(1- \nu_B \cos{[2\Delta(f_A,f_B)]}\right)\sigma^{\rm z}_B\rho^B_{-\infty}\sigma^{\rm z}_B \nonumber \\
&+&\frac{i}{2}\nu_B\sin{[2\Delta(f_A,f_B)]}\langle\sigma^{\rm z}_A\rangle_{\rho^A_{-\infty}}\left[\rho^B_{-\infty}, \sigma^{\rm z}_B \right],
\label{rhoB}
\end{eqnarray}
where $\langle\sigma^{\rm z}_A\rangle_{\rho^A_{-\infty}}\equiv {\rm tr}\left(\sigma^{\rm z}_A\rho^A_{-\infty}\right)$.

We would like to write Eq.~(\ref{rhoB}) in the form 
\begin{equation}
\rho^B = \mathcal{E}(\rho^A_{-\infty})
\label{E1}
\end{equation}
in order to define a quantum map $\mathcal{E}$ that describes the communication channel. For this purpose, we need first to fix the initial state $\rho^B_{-\infty}$ of Bob's qubit. This, however, must be done wisely; otherwise we may end up with a very poor communication channel. As $$[\sigma^{\rm z}_B, H(t)]=0,$$
where $H$ is the total Hamiltonian~(\ref{totalH}), $\sigma^{\rm z}_B$ is conserved and hence, it will not be useful to choose one of its eigenstates, $|0\rangle_B$ and $|1\rangle_B$, as the initial state for qubit B, nor measure $\rho^B$ in this basis in order to recover any information encoded by Alice. To help in the choice of $\rho^B_{-\infty}$, let us analyze the signaling between Alice and Bob. To this end, suppose that Alice encodes the message to be sent in the states ${\rho^{A}_{{}_{\!-\infty}}}_+$ and ${\rho^{A}_{{}_{\!-\infty}}}_-$ and that Bob decodes it by means of the projective measurement 
$$\left\{F^B_+\equiv |+\rangle_{{}_B}{}_{{}_B}\langle +|,F^B_-\equiv |- \rangle_{{}_B}{}_{{}_B}\langle -|\right\},$$ 
where $\sigma^{\rm x}_B|\pm\rangle_B=\pm|\pm\rangle_B$. By using Eq.~(\ref{rhoB}), the probability  $p(l|k)\equiv {\rm tr}\left(F^B_l \rho^B_k\right)$ of Bob measuring $l=\pm$ given that Alice has prepared the state ${\rho^{\!A}_{{}_{\!-\infty}}}_k$, $k=\pm$, is 
\begin{equation}
p(l|k)=\frac{1}{2}(1 + l\nu_B \Lambda_k),
\label{prob}
\end{equation}
where
\begin{equation}
\Lambda_k \!\!\equiv \! 2\Re{\left\{\!\beta_B(\!\cos{[2\Delta(f_A,f_B)]}\!- i\langle\sigma^{\rm z}_A\rangle_{\!\!{\rho^{\!A}_{{}_{\!-\infty}}}_k}\!\!\!\!\sin{[2\Delta(f_A,f_B)]})\!\right\}}
\label{prob2}
\end{equation}
and $\beta_B\equiv {}_B\langle 0|\rho^B_{-\infty}|1\rangle_B.$ Having fixed the properties of qubit $B$, given  by $f_B,$ we can see from Eq.~(\ref{prob}) and from $\Lambda_k$ in Eq.~(\ref{prob2}) that there are two terms describing the influence of qubit $A$ on qubit $B$, both of which are clearly causal [due to the appearance of $\Delta(f_A,f_B)$]. The first one, $$2\Re{\{\beta_B\cos{[2\Delta(f_A,f_B)]}\}},$$ is ``universal" in the sense that it does not depend on the state ${\rho^{A}_{-\infty}}_k$ chosen by Alice; it only depends on the properties of qubit A, described by $f_A$, and it will be there whenever its interaction with the field has been turned on.  The second term,$$-2\Re{\{\beta_Bi\langle\sigma^{\rm z}_A\rangle_{{\rho^{\!A}_{{}_{\!-\infty}}}_k}\sin{[2\Delta(f_A,f_B)]}\}},$$ however, depends explicitly on the initial state of qubit $A$ and thus, it is the one responsible for the signaling. This motivates us to choose 
\begin{equation}
\rho^B_{-\infty}\equiv |y_+\rangle_{{}_B}{}_{{}_B}\langle y_+| 
\label{rhoBfinal}
\end{equation} 
so as to maximize the signaling contribution. In Eq.~(\ref{rhoBfinal}) we have used that 
 $$ |y_+\rangle\equiv  \frac{1}{\sqrt{2}} \left(|0\rangle_B +i |1\rangle_B\right)$$ 
and hence, $\sigma^{\rm y}_B|y_+\rangle=|y_+\rangle$ and $\beta_B=-i/2$. As a result, we can write the probability in Eq.~(\ref{prob}) as 
\begin{equation}
p(l|k)=\frac{1}{2}(1 - l\nu_B \langle\sigma^{\rm z}_A\rangle_{{\rho^{A}_{-\infty}}_k}\sin{[2\Delta(f_A,f_B)]}).
\label{prob3}
\end{equation}

Having fixed the initial state for qubit $B$, the linear,  completely positive and trace-preserving (CPTP)  quantum map $\mathcal{E}$ is unambiguously defined by Eqs.~(\ref{rhoB}) and~(\ref{E1}). We can also cast $\mathcal{E}$ in terms of its Kraus decomposition~\cite{wilde}
\begin{equation}
\mathcal{E}(\rho^A_{-\infty})=\sum_{\mu=0}^3 M_\mu \rho^A_{-\infty} M_\mu^\dagger, 
\label{kraus}
\end{equation}
with Kraus operators $M_{\mu}$ given by
\begin{eqnarray}
M_0 &\equiv& \frac{1}{2}\sqrt{\frac{1-\nu_B^2}{P_e}} \; |+\rangle_{_{B}}{}_{_{A}}\langle 0|, \label{kraus1} \\
M_1 &\equiv& \frac{1}{2}\sqrt{\frac{1-\nu_B^2}{1-P_e}} \; |+\rangle_{_{B}}{}_{_{A}}\langle 1|,\\
M_2 &\equiv& \frac{i\nu_B}{2\sqrt{P_e}}\cos{[2\Delta(f_A,f_B)]} \; |+\rangle_{_{B}}{}_{_{A}}\langle 0|+ \sqrt{P_e} \; |-\rangle_{_{B}}{}_{_{A}}\langle 0|, \nonumber \\
\\
M_3 &\equiv&\frac{i\nu_B}{2\sqrt{1-P_e}}\cos{[2\Delta(f_A,f_B)]}|+\rangle_{_{B}}{}_{_{A}}\langle 1| \nonumber  \\
&+& \sqrt{1-P_e}\;|-\rangle_{_{B}}{}_{_{A}}\langle 1|,
\label{krausop}
\end{eqnarray}
where  $\sum_\mu M^\dagger_\mu M_\mu=I^A $ and 
\begin{equation}
P_e\equiv \frac{1}{2}\left(1 + \nu_B \sin{[2\Delta(f_A,f_B)]}\right).
\label{P_e}
\end{equation}
Indeed, by using Eqs.~(\ref{kraus1})-(\ref{P_e}),
\begin{eqnarray}
\!\sum_{\mu=0}^3 \!M_\mu \rho^A_{-\infty} M_\mu^\dagger\!\!\!&=&\!\!\!\frac{1}{2}\!\left(\!1\!-\!\langle\sigma^{\rm z}_A\rangle_{\rho^{A}_{-\infty}}\!\!\nu_B\sin{\!\left[2\Delta(f_A,f_B)\right]}\right)\!|+\rangle_{{}_{\!B}}\!{}_{{}_{B\!}}\langle +| \nonumber \\
&+&\!\!\!\!\frac{1}{2}\!\left(\!1\!+\!\langle\sigma^{\rm z}_A\rangle_{\rho^{A}_{-\infty}}\!\!\nu_B\sin{\!\left[2\Delta(f_A,f_B)\right]}\right)\!|-\rangle_{{}_{\!B}}\!{}_{{}_{B\!}}\langle -| \nonumber \\
&+&\!\!\left(\frac{i\nu_B}{2}\cos{\left[2\Delta(f_A,f_B)\right]}|+\rangle_{{}_{\!B}}\!{}_{{}_{B\!}}\langle -| + {\rm H.c.}\right)\!, \nonumber \\
\label{rho+basis}
\end{eqnarray}
which is $\rho^B\equiv \mathcal{E}(\rho^A_{-\infty})$ in Eq.~(\ref{rhoB}), with $\rho^B_{-\infty}=|y_+\rangle \langle y_+|,$ written in the basis $\{|+\rangle_{B},|-\rangle_B\}$. 

The quantum map in Eq.~(\ref{kraus}) gives the (nonperturbative) mathematical description of the communication channel between Alice and Bob.
\section{Channel capacities}
\label{sec:capacities}

Equipped with the map $\mathcal{E}$ given in Eq.~(\ref{kraus}), we will be able to study at which rate classical and quantum information can be reliably conveyed between Alice and Bob. We will also be able to analyze how prior entanglement shared between the two observers can aid both classical and quantum communication capacities. 

\subsection{Unassisted classical and quantum communication capacities}

Let us begin with the investigation of the unassisted channel capacities for the transmission of both classical and quantum information. In what follows, we will first review what are the classical and quantum capacities of a quantum channel (our review will follow closely the treatment given in Ref.~\cite{wilde}, where more details can be found). After that, we will compute both capacities for our communication channel.    

Suppose that Alice wants to transmit a message chosen from the set $X=\{1,\cdots, |X|\}$ to Bob, where $|X|$ indicates the cardinality of the set $X$. In order to do that, she will use the communication channel $\mathcal{E}$ described in Sec.~\ref{sec:channel}. In order to reliably convey the information through any noise channel, Alice needs to do a suitable block coding on the messages in $X$ and then make $n$ independent uses of the channel. More explicitly, to each $m\in X$ Alice associates a quantum state $\rho^{A_n}_m$ defined in the space $\mathcal{H}^{\otimes n}_A$. Then, by making $n$ independent uses of the channel $\mathcal{E}$, she transmits  $\rho^{A_n}_m$ to Bob, who receives the state $$\mathcal{E}^{\otimes n}\left(\rho^{A_n}_m\right)$$
defined in $\mathcal{H}^{\otimes n}_B$. To decode the message, Bob chooses a POVM $\{ F^{B_n}_m | m\in Y\}$, $|Y|\geq|X|$. Hence, the probability that Bob correctly determines  the message sent by Alice is given by 
\begin{equation}
P\left(Y=m|X=m\right)={\rm tr}\left[F^{B_n}_m\mathcal{E}^{\otimes n}\left(\rho^{A_n}_m\right)\right]
\end{equation}
and the error probability is 
\begin{eqnarray}
p_{\rm err}(m)&\equiv& 1- P\left(Y=m|X=m\right) \nonumber\\
&=&{\rm tr}\left[\left(I-F^{B_n}_m\right)\mathcal{E}^{\otimes n}\left(\rho^{A_n}_m\right)\right].
\end{eqnarray}
The error probability of this coding scheme is given by 
\begin{equation}
p_e^*\equiv \max_{m\in X}p_{\rm err}(m)
\end{equation}
and we say the code has error $\epsilon>0$ if 
\begin{equation}
p_e^*\leq \epsilon.
\end{equation} 
The rate $R_C$ of communication (bits per use of the channel) in this coding scheme is 
\begin{equation}
R_C\equiv \frac{1}{n}\log_2{|X|}
\end{equation} 
and we call a code with error $\epsilon$ a $(n,R_C,\epsilon)$ code for classical communication. We say that a rate $R_C$ is achievable if given any $\epsilon, \delta>0$ there exists a $(n,R_C-\delta,\epsilon)$ code for a sufficiently large $n$. 

The classical capacity of the quantum channel $\mathcal{E}$ is the supremum over all achievable rates $R_C$ and it will be denoted by $C(\mathcal{E})$. Due to the Holevo-Schumacher-Westmoreland (HSW) theorem~\cite{HSW}, $C(\mathcal{E})$ can be written as~\cite{wilde} 
\begin{equation}
C(\mathcal{E})=\lim_{k\rightarrow \infty} {\frac{1}{k}\chi(\mathcal{E}^{\otimes k})},
\label{Ccapacity}
\end{equation}
where 
\begin{equation}
\chi(\mathcal{E})\equiv \max_{\{p_m,\rho^A_m\}}{ \left\{S\left(\mathcal{E}\left[\sum_mp_m\rho^A_m\right]\right)-\sum_mp_mS\left(\mathcal{E}\left[\rho^A_m\right]\right) \right\}},
\label{chi}
\end{equation}
with $S(\rho)\equiv -{\rm tr}(\rho\log_2 \rho)$ being the von Neumann entropy of a density matrix $\rho$ and $\{p_m\}$ is a probability distribution. The quantity $\chi(\mathcal{E})$ is called the Holevo information of the channel.

Suppose now that Alice wants to transmit quantum information (qubits) to Bob by means of the communication channel $\mathcal{E}$. The quantum capacity of a quantum channel measures at which rate this can be reliably done.   We note that whenever Alice is able to reliably transmit entanglement through the channel, she will also be able to transmit an arbitrary quantum state $\rho^{T_A},$ defined on $\mathcal{H}_{T_A},$ to Bob~\cite{wilde}, since we can always think that this state arises from the entanglement of $T_A$ with some reference system $T_{A'}$. Hence, suppose that Alice and the reference system share some state $|\varphi^{T_A T_{A'}}\rangle \in \mathcal{H}_{T_A}\otimes\mathcal{H}_{T_{A'}}$, where $\mathcal{H}_{T_{A'}}$ is the Hilbert space of $T_{A'}$. Then, Alice encodes her part of the state on $\mathcal{H}^{\otimes n}_A$ by means of a CPTP map $\mathcal{C}$ ending up with the state 
\begin{equation*}
\left(\mathcal{C}\otimes I^{T_{A'}}\right)\left(\rho_\varphi^{T_AT_{A'}}\right),
\end{equation*}
where $\rho_\varphi^{T_AT_{A'}}\equiv |\varphi^{T_AT_{A'}}\rangle \langle\varphi^{T_AT_{A'}}| $ and $I ^{T_{A'}}$ is the identity operator on $\mathcal{H}_{T_{A'}}.$ She then sends her part of the encoded total state to Bob by making $n$ independent uses of the channel. The  resulting global state is 
\begin{equation}
\varpi^{B_nT_{A'}}\equiv \left(\mathcal{E}^{\otimes n}\otimes I^{T_{A'}}\right)\left(\mathcal{C}\otimes I^{T_{A'}}\left[\rho_\varphi^{T_AT_{A'}}\right]\right).
\label{BnQ}
\end{equation}
Bob decodes his share of the state~(\ref{BnQ}) by using a CPTP map $\mathcal{D}$ mapping states in $\mathcal{H}^{\otimes n}_B$ into states in $\mathcal{H}_{T_B}.$ Therefore, after Bob's decoding, the total state will be given by
\begin{equation}
\varsigma^{T_BT_{A'}}\equiv \left(\mathcal{D}\otimes I^{T_{A'}}\right)\left(\varpi^{B_nT_{A'}}\right).
\end{equation} 
The rate $R_Q$ of communication (qubits per use of the channel) in this scheme is given by 
\begin{equation}
R_Q\equiv \frac{1}{n}\log_2 {d_{T_A}},
\end{equation} 
where $d_{T_A}\equiv{\rm dim} \mathcal{H}_{T_A}.$ This coding process will be good if, for a given $\epsilon >0,$
\begin{equation}
\| \rho^{T_BT_{A'}}_\varphi -\varsigma^{T_BT_{A'}} \|_1\leq \epsilon
\end{equation}
with $$\| \mathcal{O} \|_1\equiv {\rm tr}\left(\sqrt{\mathcal{O}^\dagger \mathcal{O}}\right)$$ being the trace norm of an operator $\mathcal{O}$ and $\rho^{T_BT_{A'}}_\varphi\equiv \mathcal{I}^{T_A\rightarrow T_B}\left(\rho^{T_AT_{A'}}_\varphi\right)$. Here, $ \mathcal{I}^{T_A\rightarrow T_B}$ is the identity map between $T_A$ and $T_B$. We call the above code a $(n, R_Q,\epsilon)$ code for quantum communication.  We say that a rate $R_Q$ is achievable if, given any $\epsilon, \delta>0,$ there exists a $(n,R_Q-\delta,\epsilon)$ code for a sufficiently large $n$. 

The quantum capacity of the channel $\mathcal{E}$, denoted by $Q(\mathcal{E})$, is the  supremum over all achievable rates $R_Q$. It can be shown that this capacity can be written as~\cite{wilde,QC}
\begin{equation}
Q(\mathcal{E})=\lim_{k\rightarrow \infty} {\frac{1}{k}\mathcal{Q}(\mathcal{E}^{\otimes k})},
\label{Qcapacity}
\end{equation}
where 
\begin{equation}
\mathcal{Q}(\mathcal{E})\equiv \max_{|\varphi^{AA'}\rangle}{\left\{S(\varsigma^{B})-S(\varsigma^{BA'})\right\}},
\label{coherent}
\end{equation}
with the maximization being over all pure states $|\varphi^{AA'}\rangle\in \mathcal{H}_A\otimes \mathcal{H}_{A'}$, ${\rm dim} \mathcal{H}_{A'}={\rm dim} \mathcal{H}_A$, $\varsigma^{BA'}\equiv \left(\mathcal{E}\otimes I^{A'}\right)\left(\rho_\varphi^{AA'}\right),$ and $\varsigma^B\equiv {\rm tr}_{A'}\varsigma^{BA'}$. The quantity $\mathcal{Q}(\mathcal{E})$ is called the coherent information of the channel. It is important to stress that, whenever $C(\mathcal{E})$ and $Q(\mathcal{E})$ are nonzero, classical and quantum information, respectively, can always be transmitted from Alice to Bob with arbitrarily small error probability. However,  it takes more time (i.e. channel uses) to transmit the information through channels with lower capacity than through those with higher capacity.

In general, it is prohibitively difficult to calculate the capacities~(\ref{Ccapacity}) and~(\ref{Qcapacity}), as it may be  necessary to evaluate the Holevo and coherent information over infinite uses of the channel. Fortunately, our quantum channel $\mathcal{E}$ lies in one of the classes of channels in which the determination of both capacities is tractable, namely, the entanglement-breaking channels. A quantum channel $\mathcal{N}$ is entanglement breaking if for every  (possibly entangled) state  $|\varphi^{AA'}\rangle \in \mathcal{H}_A\otimes\mathcal{H}_{A'}$, the state $\left(\mathcal{N}\otimes I^{A'}\right)\left(\rho^{AA'}_\varphi\right)$, $\rho^{AA'}_\varphi=|\varphi^{AA'}\rangle\langle\varphi^{AA'}|$, is separable. To show that the channel $\mathcal{E}$ in Eq.~(\ref{kraus}) belongs to such a class, let us write the state $|\varphi^{AA'}\rangle$ as 
\begin{equation}
|\varphi^{AA'}\rangle=\sum_{i,i'=0}^1c_{i,i'} | i \rangle_A\otimes | i' \rangle_{A'},
\end{equation}
where   $| i \rangle_A$ and $ | i' \rangle_{A'}$ are eigenstates of $\sigma^{\rm z}_A$ and $\sigma^{\rm z}_{A'}$, respectively, and $\sum_{i,i'}|c_{ii'}|^2=1$. By using Eqs.~(\ref{kraus})-(\ref{krausop}) we have that 
\begin{equation}
\mathcal{E}\left(| i \rangle_{{}_A}{}_{{}_A}\langle i'|\right)=\delta_{i i'} \mathcal{E}\left(| i \rangle_{{}_A}{}_{{}_A}\langle i|\right) 
\end{equation}
and hence 
\begin{eqnarray}
\left(\mathcal{E}\otimes I^{A'}\right)\left(\rho^{AA'}_\varphi\right)=\sum_{i}\mathcal{E}\left(| i \rangle_{{}_A}{}_{{}_A}\langle i|\right)\otimes |\zeta_i\rangle_{{}_{A'}}{}_{{}_{A'}}\langle \zeta_i|,
\label{EBE}
\end{eqnarray}
where 
\begin{equation}
|\zeta_i\rangle_{A'} \equiv \sum_{i'}c_{i i'}|i'\rangle_{A'}.
\end{equation}
By defining the density matrices
\begin{equation}
\mathfrak{S}^B_i\equiv \mathcal{E}\left(| i \rangle_{{}_A}{}_{{}_A}\langle i|\right),
\end{equation}
and 
\begin{equation}
\tau_i^{A'}\equiv \|\zeta_i\|^{-2} |\zeta_i\rangle_{{}_{A'}}{}_{{}_{A'}}\langle\zeta_i |, 
\end{equation}
$\|\zeta_i\|^2\equiv \langle \zeta_i| \zeta_i\rangle$, we can cast Eq.~(\ref{EBE}) as 
\begin{equation}
\left(\mathcal{E}\otimes I^{A'}\right)\left(\rho^{AA'}_\varphi\right)=\sum_{i}\|\zeta_i\|^2 \mathfrak{S}^B_i\otimes \tau_i^{A'},
\label{EBE2}
\end{equation}
where we note that 
$$\sum_i\|\zeta_i\|^2=\sum_{i,i'}|c_{ii'}|^2=1.$$
The state~(\ref{EBE2}) is separable and hence, the channel $\mathcal{E}$ is entanglement breaking. It is known \cite{shorEB} that the Holevo information $\chi(\mathcal{E})$ given in Eq.~(\ref{chi}) is additive for entanglement-breaking channels, i.e., 
\begin{equation}
\chi(\mathcal{E}^{\otimes n})=n\chi\left(\mathcal{E}\right).
\end{equation}
As a consequence, the classical capacity~(\ref{Ccapacity}) can be written as $C(\mathcal{E})=\chi(\mathcal{E})$ and thus, by using the definition of $\chi(\mathcal{E})$ given in Eq.~(\ref{chi}), we can write
\begin{equation}
C(\mathcal{E})=\max_{\{p_m,\rho^A_m\}}{ \left\{S\left(\mathcal{E}\left[\sum_mp_m\rho^A_m\right]\right)-\sum_mp_mS\left(\mathcal{E}\left[\rho^A_m\right]\right) \right\}}.
\label{Ccapacity2}
\end{equation}
Contrary to the limit in Eq.~(\ref{Ccapacity}), the above maximization is a tractable problem. 

Let us now  perform the maximization in Eq.~(\ref{Ccapacity2}) and determine the classical capacity $C(\mathcal{E})$ of the communication channel. For this purpose, let $\{p_m\}$ be some probability distribution and ${\rho_{{}_{-\infty}}^A}_m$ density matrices for Alice's qubit. We will decompose each ${\rho_{{}_{-\infty}}}_m^A$ in terms of its Bloch vectors, i.e.,
\begin{equation}
{\rho_{{}_{-\infty}}^A}_m=\frac{1}{2}\left(I^A + {\bf r}_m\cdot \boldsymbol{\sigma}_A\right),
\label{rhobloch}
\end{equation}
where ${\bf r}_m\equiv (x_m,y_m,z_m)$, $\|{\bf r}_m\|^2\equiv x_m^2+y_m^2+z_m^2\leq 1$, and $\boldsymbol{\sigma}_A=(\sigma^{\rm x}_A,\sigma^{\rm y}_A,\sigma^{\rm z}_A)$. By using Eq.~(\ref{rhobloch}) in Eq.~(\ref{kraus}) [or equivalently, in Eq.~(\ref{rho+basis})], we can write the action of the quantum map $\mathcal{E}$ on ${\rho_{{}_{-\infty}}^A}_m$ and on
\begin{equation}
{\rho_{{}_{-\infty}}^A}\equiv \sum_mp_m{\rho_{{}_{-\infty}}^A}_m
\end{equation} 
as
\begin{eqnarray}
\!\mathcal{E}\left({\rho_{{}_{-\infty}}^A}_m\right)&=&\!\!\frac{1}{2}\left(1-z_m\nu_B\sin{\left[2\Delta(f_A,f_B)\right]}\right)|+\rangle_{{}_{\!B}}\!{}_{{}_{B\!}}\langle +| \nonumber \\
&+&\!\!\frac{1}{2}\left(1+z_m\nu_B\sin{\left[2\Delta(f_A,f_B)\right]}\right)|-\rangle_{{}_{\!B}}\!{}_{{}_{B\!}}\langle -| \nonumber \\
&+&\!\!\!\left(\frac{i\nu_B}{2}\cos{\left[2\Delta(f_A,f_B)\right]}|+\rangle_{{}_{\!B}}\!{}_{{}_{B\!}}\langle -| + {\rm H.c.}\!\right)
\label{rhoAm}
\end{eqnarray}
and 
\begin{eqnarray}
\!\mathcal{E}\left({\rho_{{}_{-\infty}}^A}\right)&=&\!\!\frac{1}{2}\left(1-z\nu_B\sin{\left[2\Delta(f_A,f_B)\right]}\right)|+\rangle_{{}_{\!B}}\!{}_{{}_{B\!}}\langle +| \nonumber \\
&+&\!\!\frac{1}{2}\left(1+z \nu_B\sin{\left[2\Delta(f_A,f_B)\right]}\right)|-\rangle_{{}_{\!B}}\!{}_{{}_{B\!}}\langle -| \nonumber \\
&+&\!\!\! \left(\frac{i\nu_B}{2}\cos{\left[2\Delta(f_A,f_B)\right]}|+\rangle_{{}_{\!B}}\!{}_{{}_{B\!}}\langle -| + {\rm H.c.}\right)\!,
\label{rhoAsum}
\end{eqnarray}
with $z\equiv \sum_m p_m z_m,$ respectively. By diagonalizing the density matrices in Eqs.~(\ref{rhoAm}) and~(\ref{rhoAsum}), we get their eigenvalues $p_{\mathcal{E}_m}, 1-p_{\mathcal{E}_m}$ and $p_{\mathcal{E}}, 1-p_{\mathcal{E}}$, respectively. Here, 
\begin{equation}
p_{\mathcal{E}_m}\!\equiv\!  \frac{1}{2} + \frac{\nu_B}{2}\sqrt{z_m^2\left(\sin{\left[2\Delta(f_A,f_B)\right]}\right)^2\!\!+\!\left(\cos{\left[\Delta(f_A,f_B)\right]}\right)^2} 
\label{pEm}
\end{equation}
and
\begin{equation}
p_{\mathcal{E}}\equiv  \frac{1}{2} + \frac{\nu_B}{2}\sqrt{z^2\left(\sin{\left[2\Delta(f_A,f_B)\right]}\right)^2\!\!+\!\left(\cos{\left[\Delta(f_A,f_B)\right]}\right)^2}.
\label{pE}
\end{equation}
With Eqs.~(\ref{pEm}) and~(\ref{pE}) we can write
\begin{eqnarray}
S\left(\mathcal{E}\left[{\rho_{{}_{-\infty}}^A}\right]\right)-\sum_mp_mS\left(\mathcal{E}\left[{\rho_{{}_{-\infty}}^A}_m\right]\right)\!\!&=&\!\!\!H(p_\mathcal{E}\!)\!-\!\sum_mp_mH(p_{\mathcal{E}_m}\!), \nonumber \\
\label{chiH}
\end{eqnarray}
where $H(x)\equiv -x\log_2x -(1-x)\log_2(1-x),$ $x\in [0,1]$. Now, we note that $p_{\mathcal{E}_m}\geq 1/2$,  $p_{\mathcal{E}}\geq 1/2$, and that, for $x\geq 1/2$, $H(x)$ is a monotonically decreasing function. Hence, as  
\begin{eqnarray*}
p_{\mathcal{E}_m}\leq \frac{1}{2}+ \frac{\nu_B}{2}
\label{ineq1}
\end{eqnarray*}
and 
\begin{eqnarray*}
p_{\mathcal{E}}\geq \frac{1}{2}+ \frac{\nu_B}{2}\left|\cos\left[\Delta(f_A,f_B)\right]\right|
\label{ineq1m}
\end{eqnarray*}
we conclude that 
\begin{equation}
H(p_{\mathcal{E}_m})\geq H\left(\frac{1}{2}+ \frac{\nu_B}{2}\right)
\label{ineq2}
\end{equation}
and 
\begin{equation}
H(p_{\mathcal{E}})\leq H\left(\frac{1}{2}+ \frac{\nu_B}{2}\left|\cos\left[2\Delta(f_A,f_B)\right]\right|\right).
\label{ineq3}
\end{equation}
By defining 
\begin{equation}
\chi\left(\mathcal{E}\left[\rho^A_{-\infty}\right]\right)\equiv S\left(\mathcal{E}\left[{\rho_{{}_{-\infty}}^A}\right]\right)-\sum_mp_mS\left(\mathcal{E}\left[{\rho_{{}_{-\infty}}^A}_m\right]\right)\! \!\!
\end{equation}
and using Eqs.~(\ref{ineq2}) and~(\ref{ineq3}) in Eq.~(\ref{chiH}) we end up with 
\begin{eqnarray}
\chi\left(\mathcal{E}\left[\rho^A_{-\infty}\right]\right)&\leq &H\left(\frac{1}{2}+ \frac{\nu_B}{2}\left|\cos\left[2\Delta(f_A,f_B)\right]\right|\right) \nonumber \\
&-& H\left(\frac{1}{2}+ \frac{\nu_B}{2}\right).
\label{boundchi}
\end{eqnarray}
The upper bound in Eq.~(\ref{boundchi}) can be reached if one chooses, e.g., a probability distribution $\{p_m\}$ with $p_1=p_2=1/2$ (and thus $p_{m>2}=0$) and the Bloch vectors ${\bf r}_1=(0,0,1),$ ${\bf r}_2=(0,0,-1),$ and ${\bf r}_{m>2}=(x_m,y_m,0)$  in Eq.~(\ref{rhobloch}). As a consequence, we can write the classical capacity in Eq.~(\ref{Ccapacity2}) as 
\begin{equation}
C\left(\mathcal{E}\right)=H\left(\frac{1}{2}+ \frac{\nu_B}{2}\left|\cos\left[2\Delta(f_A,f_B)\right]\right|\right) - H\left(\frac{1}{2}+ \frac{\nu_B}{2}\right).
\label{Ccapacityfinal}
\end{equation}
 Whenever Alice and Bob interacts with field in causally disconnected regions, the supports of the functions $f_A$ and $f_B$  are spacelike separated  and thus $\Delta(f_A,f_B)=0$. Then, by using Eq.~(\ref{Ccapacityfinal}), $C(\mathcal{E})=0$ and hence, it is impossible for Alice and Bob to communicate. If, however, Bob's interaction with the field is in the future of Alice's, $C(\mathcal{E})$ will be nonzero and, as a consequence, Alice can always reliably convey her message to Bob.

Now, the fact that the channel $\mathcal{E}$ is entanglement breaking also allow us to compute its quantum capacity $Q(\mathcal{E})$, given in Eq.~(\ref{Qcapacity}).  As shown, e.g., in~\cite{holevoEB}, the quantum capacity of an entanglement-breaking channel (actually, of any antidegradable channel) is 
\begin{equation}
Q(\mathcal{E})=0.
\label{Qcapacityfinal}
\end{equation}
Therefore, Alice cannot convey quantum information to Bob by making use of the channel $\mathcal{E}.$

\subsection{Entanglement-assisted classical and quantum communication capacities}

We have seen in the last subsection that Alice can always reliably convey classical information to Bob, by using the quantum channel $\mathcal{E}$, whenever their interactions with the field are causally connected. In contrast, Alice is not able to transmit quantum information to Bob under any circumstance. We would like to analyze now how the capacities to send both classical and quantum information are affected when Alice and Bob initially have access to an unlimited supply of entanglement.  For this purpose, let us first describe the protocol for this entanglement-assisted quantum communication. Again, we follow the treatment given in Ref.~\cite{wilde}, were more details can be found. 

Suppose, for simplicity, that Alice and Bob share a maximally entangled state 
$$|\Phi^{T_AT_{B'}}\rangle=\frac{1}{\sqrt{d}}\sum_{i=0}^{d-1}|i\rangle_{T_A}\otimes|i\rangle_{T_{B'}},$$
defined on $\mathcal{H}_{T_A}\otimes\mathcal{H}_{T_{B'}}$, where $\{|i\rangle_{T_a}\}, a=A,B',$ is an orthonormal set of vectors on $\mathcal{H}_{T_a}$, and $d$ can be as large as they need. Alice wants to transmit a message chosen from the set $X=\{1,\cdots, |X|\}$ to Bob by means of the communication channel $\mathcal{E}$. To this end, she associates to each $m\in X$ a CPTP map $\mathcal{C}_m$ taking  states defined on $\mathcal{H}_{T_A}$ into states defined on $\mathcal{H}^{\otimes n}_A$. If she chooses to send the message $m$ to Bob, she applies $\mathcal{C}_m$ on her half of the entangled state $|\Phi^{T_AT_{B'}}\rangle$. The total state then becomes $$\left(\mathcal{C}_m\otimes I^{T_{B'}}\right)(\rho^{T_AT_{B'}}_\Phi),$$
where $\rho^{T_AT_{B'}}_\Phi \equiv |\Phi^{T_AT_{B'}}\rangle\langle \Phi^{T_AT_{B'}}|$. After the encoding, Alice  sends her share of the total state to Bob by making n independent uses of the channel $\mathcal{E}.$ After receiving Alice's part of the entangled state, Bob will be in possession of the state 
\begin{equation}
\varpi_m^{B_nT_{B'}}\equiv \left(\mathcal{E}^{\otimes n}\otimes I^{T_{B'}}\right)\left(\mathcal{C}_m\otimes I^{T_{B'}}\left[\rho^{T_AT_{B'}}_\Phi\right]\right),
\label{BnB'}
\end{equation}
defined in $\mathcal{H}^{\otimes n}_B\otimes \mathcal{H}_{T_{B'}}$. To decode the message, he chooses a POVM $\{ F^{B_nT_{B'}}_m | m\in Y\}$, $|Y|\geq|X|$ to perform a measurement on the total state~(\ref{BnB'}). Hence, the probability that Bob correctly determines  the message sent by Alice is given by 
\begin{equation}
P\left(Y=m|X=m\right)={\rm tr}\left[F^{B_nT_{B'}}_m\varpi_m^{B_nT_{B'}}\right]
\end{equation}
and the error probability is 
\begin{eqnarray}
p_{\rm err}(m)&\equiv& 1- P\left(Y=m|X=m\right) \nonumber\\
&=&{\rm tr}\left[(I-F^{B_nT_{B'}}_m)\varpi_m^{B_nT_{B'}}\right].
\end{eqnarray}
The error probability of this coding scheme is given by 
\begin{equation}
p_e^*\equiv \max_{m\in X}p_{\rm err}(m)
\end{equation}
and we say that the code has error $\epsilon>0$ if 
\begin{equation}
p_e^*\leq \epsilon.
\end{equation} 
The rate $R_{ea}$ of communication (bits per use of the channel) in this coding scheme is 
\begin{equation}
R_{ea}\equiv \frac{1}{n}\log_2{|X|}
\end{equation} 
and we call a code with error $\epsilon$ an $(n,R_{ea},\epsilon)$ code for entanglement-assisted classical communication. We say that a rate $R_{ea}$ is achievable if, given any $\epsilon, \delta>0$, there exists a $(n,R_{ea}-\delta,\epsilon)$ code for a sufficiently large $n$. 

The entanglement-assisted classical capacity of the channel is the supremum over all achievable rates $R_{ea}$ and it will be denoted by $C_{ea}\left(\mathcal{E}\right)$. As proved in~\cite{BSST02}, this capacity can be written as
 \begin{equation}
C_{ea}(\mathcal{E})=\max_{|\varphi^{AB'}\rangle}{ \left\{S(\varsigma^B) + S(\varsigma^{B'})-S(\varsigma^{BB'})  \right\}},
\label{CcapacityEA}
\end{equation}
where $\varsigma^{BB'}\equiv \left(\mathcal{E}\otimes I^{B'}\right)\left(\rho^{AB'}_\varphi\right),$ with $\rho^{AB'}_\varphi \equiv |\varphi^{AB'}\rangle\langle \varphi^{AB'}|,$ $\varsigma^B={\rm tr}_{B'}\varsigma^{BB'},$ $\varsigma^{B'}={\rm tr}_{B}\varsigma^{BB'},$ and the maximization is over all pure states  $|\varphi^{AB'}\rangle \in \mathcal{H}_A\otimes\mathcal{H}_{B'}.$ The right-hand side of Eq.~(\ref{CcapacityEA}) is called the mutual information of the channel $\mathcal{E}$. 

The entanglement-assisted quantum capacity of the channel, denoted by $Q_{ea}(\mathcal{E})$, is the maximum rate at which qubits can be reliably sent through the channel when Alice and Bob share an unlimited amount of entanglement.  By using teleportation in conjunction with superdense coding, it can be shown that~\cite{wilde} 
\begin{equation}
Q_{ea}(\mathcal{E})=\frac{1}{2}C_{ea}(\mathcal{E}).
\label{QcapacityEA}
\end{equation}
We note that the expressions~(\ref{CcapacityEA}) and~(\ref{QcapacityEA}) for the entanglement-assisted classical and quantum capacity, respectively, are calculable for any quantum channel, in contrast with the expressions~(\ref{Ccapacity}) and~(\ref{Qcapacity}) for the unassisted capacities. 

Let us begin by computing the entanglement-assisted classical capacity~(\ref{CcapacityEA}), since its calculation will also determine the entanglement-assisted quantum capacity~(\ref{QcapacityEA}). To this end, let us first note that we can rewrite the Kraus operators in Eqs.~(\ref{kraus1})-(\ref{krausop}) in the following form:

\begin{eqnarray}
M_0 \!\!&=\!& \frac{1}{2}\sqrt{\frac{1-\nu_B^2}{P_e}} \; |+\rangle_{_{B}}{}_{_{A}}\langle 0|, \label{kraus1b} \\
M_1 \!\!&=&\! \frac{1}{2}\sqrt{\frac{1-\nu_B^2}{1-P_e}} \; |+\rangle_{_{B}}{}_{_{A}}\langle 1|,\\
M_2 \!\!&=&\!\!\sqrt{P_e+\frac{\nu_B^2}{4P_e}(\cos{[2\Delta(f_A,f_B)]})^2}|\mathfrak{n}_0\rangle_{_{B}}{}_{_{A}}\langle 0|, \\
M_3 \!\!&=&\! \!\sqrt{1-P_e+\frac{\nu_B^2}{4(1-P_e)}(\cos{[2\Delta(f_A,f_B)]})^2}|\mathfrak{n}_1\rangle_{_{B}}{}_{_{A}}\langle 1|, \nonumber \\ \label{krausopb}
\end{eqnarray}
where
\begin{eqnarray}
|\mathfrak{n}_0\rangle&\equiv& \frac{i\nu_B\cos{[2\Delta(f_A,f_B)]}}{\sqrt{4P_e^2+\nu_B^2(\cos{[2\Delta(f_A,f_B)]})^2}} |+\rangle_{_{B}} \nonumber \\
&+& \frac{2 P_e}{\sqrt{4P_e^2+\nu_B^2(\cos{[2\Delta(f_A,f_B)]})^2}} |-\rangle_{_{B}}
\end{eqnarray}
and 
\begin{eqnarray}
|\mathfrak{n}_1\rangle &\equiv &\frac{i\nu_B\cos{[2\Delta(f_A,f_B)]}}{\sqrt{4(1-P_e)^2+\nu_B^2 (\cos{[2\Delta(f_A,f_B)]})^2}}|+\rangle_{_{B}}\nonumber \\
&+& \frac{2(1-P_e)}{\sqrt{4(1-P_e)^2+\nu_B^2 (\cos{[2\Delta(f_A,f_B)]})^2}}|-\rangle_{_{B}}.\nonumber \\
\label{n1}
\end{eqnarray}
Then, by using Eqs.~(\ref{kraus1b})-(\ref{krausopb}) in Eq.~(\ref{kraus}) we can write
\begin{equation}
\mathcal{E}\left(\rho^A_{-\infty}\right)={}_A\langle 0|\rho^A_{-\infty}|0\rangle_A \mathfrak{S}^B_0 + {}_A\langle 1|\rho^A_{-\infty}|1\rangle_A \mathfrak{S}^B_1, 
\label{cqmap}
\end{equation}
where
\begin{eqnarray}
\mathfrak{S}^B_0&=&\left\{P_e+\frac{\nu_B^2}{4P_e}(\cos{[2\Delta(f_A,f_B)]})^2\right\}|\mathfrak{n}_0\rangle_{{}_B}{}_{{}_B}\langle  \mathfrak{n}_0| \nonumber \\
&+& \frac{1 -\nu_B^2}{4 P_e}|+\rangle_{{}_B}{}_{{}_B}\langle +| 
\label{w0}
\end{eqnarray}
and 
\begin{eqnarray}
\mathfrak{S}^B_1\!\!&=&\!\!\!\!\left\{1-P_e+\frac{\nu_B^2}{4(1-P_e)}(\cos{[2\Delta(f_A,f_B)]})^2\right\}|\mathfrak{n}_1\rangle_{{}_B}{}_{{}_B}\langle \mathfrak{n}_1| \nonumber \\
&+& \frac{1 -\nu_B^2}{4 (1-P_e)}|+\rangle_{{}_B}{}_{{}_B}\langle +|,  
\label{w1}
\end{eqnarray}
and we note that, by using Eqs.~(\ref{P_e}),~(\ref{w0}) and~(\ref{w1}), ${\rm tr}(\mathfrak{S}^B_i)=1, i=0,1.$ A channel of the form~(\ref{cqmap}) is called a classical-quantum (c-q) channel. 
 
The definition of the unassisted and entanglement-assisted classical capacities $C(\mathcal{E})$ and $C_{ea}(\mathcal{E})$, respectively, implies that 
\begin{equation}
C_{ea}(\mathcal{E})\geq C(\mathcal{E}).
\end{equation}
A simple example where this inequality is manifest is the superdense coding for noiseless channels~\cite{wilde}, where it is possible for Alice to send two bits of classical information to Bob by using the prior entanglement that they shared. Notwithstanding this, it was proven in Ref.~\cite{shirokov} that, for c-q channels, $C_{ea}\!(\mathcal{E})=C(\mathcal{E})$ and hence, by using Eq.~(\ref{Ccapacityfinal})
\begin{equation}
C_{ea}(\mathcal{E})=H\left(\frac{1}{2}+ \frac{\nu_B}{2}\left|\cos\left[2\Delta(f_A,f_B)\right]\right|\right)- H\left(\frac{1}{2}+ \frac{\nu_B}{2}\right).
\label{EACcapacity}
\end{equation}
Thus, we conclude that it is not worth using a valuable resource such as entanglement in order to try to increase the classical capacity of the channel $\mathcal{E}$. Even if Alice and Bob initially share an unlimited amount of entanglement, the capacity to send classical information will always be the same as if no prior entanglement is shared. This picture changes dramatically in the case of quantum communication. In contrast to the unassisted case, where the quantum capacity vanishes, when Alice and Bob initially share entanglement, Alice can reliably send quantum information to Bob at a rate 
\begin{equation}
Q_{ea}(\mathcal{E})\!\!=\!\!\frac{1}{2}\left[\!H\left(\frac{1}{2}+ \frac{\nu_B}{2}\left|\cos\left[2\Delta(f_A,f_B)\!\right]\right|\right) \!-\! H\left(\frac{1}{2}+ \frac{\nu_B}{2}\right)\right].
 \label{EAQcapacity}
 \end{equation}
Equations~(\ref{EACcapacity}) and~(\ref{EAQcapacity}) clearly demonstrate the causality in the communication of both classical and quantum information in the entanglement-assisted case. We can see that whenever Bob's qubit interaction with the field is causally disconnected from the interaction of Alice's qubit with the field,  $C_{ea}(\mathcal{E})=0$ and $Q_{ea}(\mathcal{E})=0$ as it should be.

\section{Conclusions}
\label{sec:finalremarks}

 In this paper, we have analyzed communication of both classical and quantum information using a bosonic quantum field as a communication channel. The model we have considered encompasses several aspects desirable in relativistic quantum communication scenarios, namely: {\bf (1)} it is valid in arbitrary globally hyperbolic spacetimes, {\bf (2)} no choice of representation of the CCR needs to be made, {\bf (3)} both sender and receiver are allowed arbitrary motions, and {\bf (4)} both sender and receiver can interact with the quantum field only in a bounded region of the spacetime. In addition, the model allowed us to trace out the field's degrees of freedom in an exact manner and hence, we could determine the quantum channel between the sender and receiver nonpertubatively.

To determine the communication channel, we have considered that the sender, Alice, prepares some input state $\rho^A_{-\infty}$ for her qubit and switches on its interaction with the field for a finite time $\Delta t_A$. After that, the receiver, Bob,  switches on his qubit interaction with the field for a finite time  $\Delta t_B$ in order to measure the information imprinted by Alice on the field's state. The initial state, $\rho^B_{-\infty}$, of Bob's qubit was chosen to be in one of the eigenstates of $\sigma^{\rm y}_B$ in order to maximize the signaling between Alice and Bob. We have supposed that the field was in some arbitrary quasifree state $\omega_\mu$. After tracing out the field degrees of freedom, we have obtained the quantum map $\mathcal{E}$ that describes the communication channel. We also have cast $\mathcal{E}$ terms of its Kraus decomposition and proved that it is an entanglement-breaking channel.

In possession of the quantum channel, we have studied the maximal rate at which both classical and quantum information can be sent through it with an arbitrary small error probability in the reception. These maximum rates, also called the classical and quantum capacities of the channel, were analyzed in two situations. The first one is when Alice and Bob do not initially share entanglement (the unassisted case). The second one is when they  share an unlimited amount of entanglement before communicating (the entanglement-assisted case). For both the unassisted and entanglement-assisted cases, the quantum and classical capacities vanish whenever Alice and Bob are spacelike separated and try to communicate.  Hence, causality is manifest in this communication model. In the unassisted case, we have seen that the classical capacity is nonvanishing when Alice and Bob are causally connected and thus, Alice is able to reliably convey a classical message to Bob. On the other hand, the quantum capacity is identically zero and hence, it is impossible for Alice to reliably send qubits to Bob. In the entanglement-assisted case, the prior entanglement does not increase Alice's capacity to send classical information to Bob when compared to the unassisted one. For the transmission of quantum information,  however, the initial entanglement shared between Alice and Bob enables Alice to reliably convey quantum information to Bob, in sharp contrast to the unassisted case. This happens because the entanglement-assisted quantum capacity is nonzero whenever Alice and Bob are causally connected when they try to communicate.

\acknowledgments

This work was partially supported by S\~ao Paulo Research Foundation 
(FAPESP) under Grant No. 2014/26307-8.


\begin{thebibliography}{}
\bibitem{footnote1} To be precise, we need to take $\mathcal{S}^\mathbb{C}$ as the space of complex solutions of Eq.~(\ref{KG}) with compact support initial data on $\Sigma_t$. Then, to define the splitting $\mathcal{S}^\mathbb{C}\simeq\mathcal{H}\oplus\overline{\mathcal{H}}$ with the desired properties, we need first to suitably ``enlarge" $\mathcal{S}^\mathbb{C}$. For details see~\cite{wald94}.
\bibitem{footnote2} Actually, it can be shown~\cite{simon} that $\mathcal{W}(\mathcal{M})$ can be made a $C^*$-algebra by a suitable completion process. It is this $C^*$-algebra that defines $\mathcal{W}(\mathcal{M})$.
\end{thebibliography}
\end{document}